\documentclass[fp,twocolumn]{jpsj3}

\usepackage{graphicx}
\usepackage{txfonts}

\title{Magnetic properties and superconductivity in the two-dimensional repulsive Hubbard model}

\author{Alexei Sherman}

\inst{Institute of Physics, University of Tartu, W. Ostwaldi Str 1, 50411 Tartu, Estonia}

\abst{
A new method for estimating the parameter ensuring the fulfillment of the Mermin-Wagner theorem in the strong coupling diagram technique (SCDT) for the two-dimensional Hubbard model is suggested. With the precise parameter value, calculated magnetic quantities are in good agreement with the results of numeric and optical-lattice experiments. Obtained spin and charge vertices are used for investigating superconductivity in the $t$-$U$ and $t$-$t'$-$t''$-$U$ Hubbard models in the regime of strong correlations. We found no superconducting transition in the $t$-$U$ model. In the $t$-$t'$-$t''$-$U$ model, the transition occurs for the singlet $d_{x^2-y^2}$ pairing at $T_c\approx0.016t$. The difference between the two models is in the renormalized hopping describing electron motion in SCDT. In the $t$-$U$ model, it vanishes for momenta $(\pi,0)$, $(0,\pi)$ of extrema of the $d$-wave order parameter, while the hopping is finite in the $t$-$t'$-$t''$-$U$ model. In the one-band model, there are optimal values of $t'$ and $t''$ ensuring the highest $T_c$.
}

\begin{document}
\maketitle

\section{Introduction}
The strong coupling diagram technique (SCDT) was designed for models of strongly correlated electrons. It is based on the regular series expansion in powers of hopping constants around the atomic limit. The idea of such an expansion was suggested by Hubbard \cite{Hubbard} and developed in a number of later works. \cite{Zaitsev,Vladimir,Metzner,Pairault,Sherman18,Sherman19} The approach is especially useful for the strong-coupling case, which, in particular, manifests itself in its ability to describe the Mott metal-insulator transition in the one-band repulsive Hubbard model.\cite{Zaitsev} In this regard, the SCDT compares favorably with the weak coupling diagram technique with the series expansion around the non-interacting limit. The latter approach faces difficulties in the description of the transition and Hubbard bands.\cite{Vilk}

In the repulsive one-band Hubbard model on a square lattice, the SCDT describes the transition to the long-range antiferromagnetic order (LRAFO) in the spin subsystem. However, in contradiction with the Mermin-Wagner theorem,\cite{Mermin} the transition temperature is nonvanishing. To remedy this defect, we suggested\cite{Sherman19} introducing the parameter $\zeta$ in the second-order cumulant playing the role of the irreducible four-leg vertex in the SCDT. The parameter decreased spin fluctuations and shifted the transition temperature to zero. Actually, the method of Ref.~\citen{Sherman19} determined $\zeta$ for half-filling and zero temperature -- conditions for the most pronounced spin fluctuations. Nevertheless, densities of electron states, spectral functions, and double occupancies obtained with this $\zeta$ were in good agreement with exact-diagonalization and Monte Carlo results in wide ranges of temperature and doping.\cite{Sherman18,Sherman19} However, for quantities derived from the calculated spin susceptibility, the agreement with outcomes of numerical and optical-lattice experiments is less satisfactory. The cause of this fault is insufficient accuracy in determining $\zeta$ and high sensitivity of the staggered susceptibility to this parameter. Hence, to improve the agreement, we should use another way for a precise evaluation of the $\zeta$.

In this work, we determine $\zeta$ by equalling two expressions for the squared site spin. This quantity is connected by a simple relation with the double occupancy, which can be calculated from the one-particle Green's function.\cite{Vilk} The second way for calculating the squared site spin is from the spin susceptibility. The parameter $\zeta$, which equalizes these two expressions, depends on doping and temperature and, for half-filling and zero temperature, acquires values found earlier.\cite{Sherman19} Values of the staggered susceptibility, squared site spin, spin structure factor obtained with such $\zeta$ turn out to be in good agreement with results of Monte Carlo simulation, numerical linked-cluster expansions (NLCE), and experiments with ultracold atoms in two-dimensional (2D) optical lattices. Besides, the obtained frequency dependence of the spin susceptibility demonstrates a correct asymptotic behavior.

In Ref.~\citen{Sherman21}, the possibility of superconductivity in the Hubbard model in the regime of strong correlations was studied using the SCDT. The range of parameters was chosen outside the regions of negative electron compressibility to avoid its influence on the solutions of the Eliashberg equation.\cite{Eliashberg} Both singlet and triplet order parameters corresponding to all one-dimensional representations of the $D_4$ point group were considered. With $\zeta$ defined in Ref.~\citen{Sherman19}, no superconducting transition was found in the $t$-$U$ and $t$-$t'$-$t''$-$U$ models. For values of this parameter calculated in the present work, at low temperatures, the spin vertex increases considerably compared with the previous estimates.
This vertex plays an important role in the singlet $d_{x^2-y^2}$ pairing. Nevertheless, with new $\zeta$ values, we found no superconducting transition in the $t$-$U$ model. However, for the $t$-$t'$-$t''$-$U$ model, the transition to the superconducting state with the $d_{x^2-y^2}$ pairing occurs at $T_c\approx0.016t\approx37$~K. The difference between the two models is in the renormalized electron hopping $\theta$ entering, together with the vertex, the matrix of the Eliashberg equation. In the $t$-$U$ model, $\theta$ vanishes for momenta $(\pi,0)$, $(0,\pi)$, in which extrema of the $d$-wave order parameter are located. This peculiarity significantly reduces the eigenvalue of the equation. In the $t$-$t'$-$t''$-$U$ model, $\theta$ is finite in these points, and the eigenvalue is larger. Due to frustrations introduced by distant electron hopping into the spin subsystem, there are optimal values of $t'$ and $t''$, ensuring the highest $T_c$.

The paper is organized as follows. In Section~2, the main SCDT equations and the equation for calculating $\zeta$ are given, and methods of their solutions are briefly discussed. Calculated squared site spin, staggered susceptibility, spin structure factor, and double occupancy are compared with results of numeric and optical-lattice experiments in Section~3. Here the asymptotic behavior of the frequency dependence of the spin susceptibility is also considered. The solution of the Eliashberg equation is discussed in Section~4. The last section is devoted to concluding remarks.

\section{Model and SCDT method}
The Hamiltonian of the fermionic Hubbard model on a square lattice reads
\begin{equation}\label{Hamiltonian}
H=\sum_{\bf ll'\sigma}t_{\bf ll'}a^\dagger_{\bf l'\sigma}a_{\bf l\sigma}
+\frac{U}{2}\sum_{\bf l\sigma}n_{\bf l\sigma}n_{\bf l,-\sigma},
\end{equation}
where 2D vectors ${\bf l}$ and ${\bf l'}$ label sites of a lattice, $\sigma=\uparrow,\downarrow$ is the spin projection, $a^\dagger_{\bf l\sigma}$ and $a_{\bf l\sigma}$ are electron creation and annihilation operators, $t_{\bf ll'}$ is the hopping constant and $n_{\bf l\sigma}=a^\dagger_{\bf l\sigma}a_{\bf l\sigma}$ is the number operator. In this work, two cases of hopping constants are considered. In one of them, only the integral between nearest-neighbor sites $t$ is nonvanishing. In the second case, the integrals between second $t'=-0.3t$ and third $t''=0.2t$ neighbors are taken into account also. These values of $t'$ and $t''$ were suggested by band-structure calculations.\cite{Andersen}

In the present work, we consider the following three Green's functions:
\begin{eqnarray}
&&G({\bf l'},\tau';{\bf l},\tau)=\langle{\cal T}\bar{a}_{\bf l'\sigma}(\tau')a_{\bf l\sigma}(\tau)\rangle, \label{Gf} \\
&&\chi^{\rm sp}({\bf l'},\tau';{\bf l},\tau)=\langle{\cal T}\bar{a}_{\bf l'\uparrow}(\tau')a_{\bf l'\downarrow}(\tau')\bar{a}_{\bf l\downarrow}(\tau)a_{\bf l\uparrow}(\tau)\rangle, \label{chisp}\\
&&\chi^{\rm sc}=\frac{1}{N}\int_{0}^{\beta}\sum_{\bf lm}\sum_{\bf l'm'}\phi_{\bf lm}\phi^*_{\bf l'm'} \nonumber\\
&&\quad\times\langle{\cal T}a_{\bf m\downarrow}(\tau)a_{\bf l\uparrow}(\tau)\bar{a}_{\bf l'\uparrow}\bar{a}_{\bf m'\downarrow}\rangle, \label{chisc}
\end{eqnarray}
the one-particle Green's function, spin and zero-frequency homogeneous superconducting susceptibilities, respective\-ly. In the above formulas, the statistical averaging denoted by the angular brackets and time dependencies
$$\bar{a}_{\bf l\sigma}(t)=\exp{({\cal H}\tau)}a^\dagger_{\bf l\sigma}\exp{(-{\cal H}\tau})$$
are determined by the operator ${\cal H}=H-\mu\sum_{\bf l\sigma}n_{\bf l\sigma}$ with the chemical potential $\mu$, ${\cal T}$ is the chronological operator, $N$ is the number of sites, $\beta=1/T$ is the inverse temperature, and $\phi_{\bf lm}$ is the pairing function.

In SCDT,\cite{Vladimir,Metzner,Pairault,Sherman18,Sherman19} such Green's functions are calculated using series expansions in powers of hopping constants. This approach is well suited for the case of strong correlations, $U\gg|t_{\bf ll'}|$. Terms of the SCDT series are products of hopping constants and on-site cumulants of creation and annihilation operators. These terms can be visualized by depicting $t_{\bf ll'}$ as a directed line connecting the sites and an on-site cumulant as a circle. The number of lines outgoing from and incoming to the circle is equal to the number of creation and annihilation operators in the cumulant. As for the weak coupling diagram technique,\cite{Abrikosov} the linked-cluster theorem is valid and partial summations are allowed in the SCDT. The notion of the one-particle irreducible diagram can be introduced in this diagram technique also. It is a two-leg diagram, which cannot be divided into two disconnected parts by cutting a hopping line. If we denote the sum of all such diagrams by the symbol $K$, the Fourier transform of Green's function (\ref{Gf}) can be written as
\begin{equation}\label{Larkin}
G({\bf k},j)=\big\{[K({\bf k},j)]^{-1}-t_{\bf k}\big\}^{-1},
\end{equation}
where {\bf k} is the 2D wave vector, $j$ is an integer defining the Matsubara frequency $\omega_j=(2j-1)\pi T$, and $t_{\bf k}$ is the Fourier transform of $t_{\bf ll'}$.

In the same manner, notions of particle-hole and particle-particle irreducible four-leg diagrams can be introduced. These diagrams cannot be divided into two disconnected parts by cutting a pair of oppositely directed and unidirectional hopping lines, respectively. In the following consideration, we use the ladder approximation, in which reducible four-leg vertices are represented by infinite sums of ladder diagrams of all lengths. This approach allows one to describe interactions of electrons with spin and charge fluctuations of all lengths in an infinite crystal. In our consideration, ladders are constructed from cumulants of the first $C^{(1)}$ and second $C^{(2)}$ orders and renormalized hopping lines
\begin{equation}\label{tta}
\theta({\bf k},j)=t_{\bf k}+t_{\bf k}^2G({\bf k},j).
\end{equation}
This renormalization is the result of the partial summation -- insertions of all possible sequences of two-leg irreducible diagrams in hopping lines, which is possible for all internal lines.

\begin{figure}[t]
\vspace*{9ex}
\centering
\includegraphics[width=0.99\columnwidth]{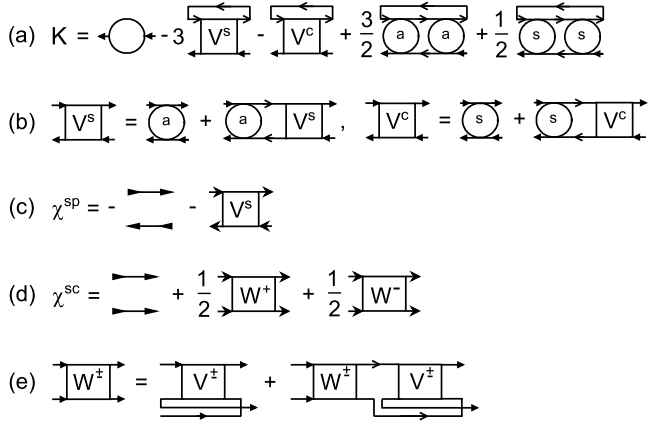}
\caption{The diagrammatic representation of main equations. Circles represent cumulants, circles with letters a and s are antisymmetrized and symmetrized second-order cumulants. Lines with two arrows at the ends are electron Green's functions. To distinguish endpoints of vertices from arrowed lines picturing $\theta$ and $\Pi$, different arrowheads are used.} \label{Fig1}
\end{figure}

In this approximation, the irreducible part $K({\bf k},j)$ in Eq.~(\ref{Larkin}) reads\cite{Sherman18}
\begin{eqnarray}
&&K({\bf k},j)=C^{(1)}(j)-\frac{T}{2N}\sum_{{\bf k'}j}\theta({\bf k'},j')[3V^s_{\bf k-k'}(j,j,j',j')\nonumber\\
&&\quad+V^c_{\bf k-k'}(j,j,j',j')]\nonumber\\
&&\quad+\frac{T^2}{4N^2}\sum_{{\bf k'}j'\nu}\theta({\bf k'},j'){\cal T}_{\bf k-k'}(j+\nu,j'+\nu) \nonumber\\
&&\quad[3C^{(2a)}(j,j+\nu,j'+\nu,j')C^{(2a)}(j+\nu,j,j',j'+\nu)\nonumber\\
&&\quad+C^{(2s)}(j,j+\nu,j'+\nu,j')C^{(2s)}(j+\nu,j,j',j'+\nu)],\label{K}
\end{eqnarray}
where ${\cal T}_{\bf k}(j,j')=N^{-1}\sum_{\bf k'}\theta({\bf k'+k},j)\theta({\bf k'},j')$, $\nu$ is an integer, $V^s$ is an infinite sum of ladder diagrams described by the following Bethe-Salpeter equation (BSE):
\begin{eqnarray}
&&V^s_{\bf k}(j+\nu,j,j',j'+\nu)=C^{(2a)}(j+\nu,j,j',j'+\nu)\nonumber\\
&&\quad+T\sum_{\nu'}C^{(2a)}(j+\nu,j+\nu',j'+\nu',j'+\nu)\nonumber\\
&&\quad\times{\cal T}_{\bf k}(j+\nu',j'+\nu')V^s_{\bf k}(j+\nu',j,j',j'+\nu'),\label{Vs}
\end{eqnarray}
and $V^c$ is another infinite sum of ladder diagrams de\-scri\-bed by a similar BSE, in which $C^{(2a)}$ is substituted with $C^{(2s)}$. These two quantities are second-order cumulants, which are antisymmetrized and symmetrized over spin indices,
\begin{eqnarray*}
&&C^{(2a)}(j+\nu,j,j',j'+\nu)\\
&&\quad=\sum_{\sigma'}\sigma\sigma'
C^{(2)}(j+\nu,\sigma';j,\sigma;j,\sigma;j'+\nu,\sigma'),\\
&&C^{(2s)}(j+\nu,j,j',j'+\nu)\\
&&\quad=\sum_{\sigma'}C^{(2)}(j+\nu,\sigma';j,\sigma;j,\sigma;j'+\nu,\sigma').
\end{eqnarray*}
The quantities $V^s$ and $V^c$ are sets of ladder diagrams con\-tri\-bu\-ting to spin and charge susceptibilities (see equation~(\ref{sp})), which is indicated by their superscripts. Diagrams corresponding to the Eq.~(\ref{K}) are shown in Fig.~\ref{Fig1}(a), BSEs for $V^s$ and $V^c$ are depicted in Fig.~\ref{Fig1}(b).

For the one-band Hubbard model, the first- and second-order on-site cumulants read
\begin{eqnarray}
&&C^{(1)}(j)=Z^{-1}\Big[\Big({\rm e}^{-\beta E_1}+{\rm e}^{-\beta E_0}\Big)g_{01}(j) \nonumber\\
&&\quad+\Big({\rm e}^{-\beta E_1}+{\rm e}^{-\beta E_2}\Big)g_{12}(j)\Big],\label{C1}\\
&&C^{(2)}(j+\nu,\sigma';j,\sigma;j',\sigma;j'+\nu,\sigma')\nonumber\\
&&\quad=\Big[Z^{-1}{\rm e}^{-\beta E_1}(\beta\delta_{\nu0}-\beta'\delta_{jj'}\delta_{\sigma\sigma'})\nonumber\\
&&\quad-Z^{-2}\Big({\rm e}^{-\beta(E_0+E_2)}
-{\rm e}^{-2\beta E_1}\Big)(\beta'\delta_{jj'} -\beta\delta_{\nu0}\delta_{\sigma\sigma'})\Big] \nonumber\\
&&\quad\times F(j)F(j'+\nu)\nonumber\\
&&\quad+Z^{-1}{\rm e}^{-\beta E_0}\delta_{\sigma,-\sigma'}Ug_{01}(j)g_{01}(j'+\nu) \nonumber\\
&&\quad\times g_{02}(\omega_j+\omega_{j'+\nu})[g_{01}(j')+g_{01}(j+\nu)]\nonumber\\
&&\quad+Z^{-1}{\rm e}^{-\beta E_2}\delta_{\sigma,-\sigma'}Ug_{12}(j)g_{12}(j'+\nu) \nonumber\\
&&\quad\times g_{02}(\omega_j+\omega_{j'+\nu})[g_{12}(j')+g_{12}(j+\nu)]\nonumber\\
&&\quad-Z^{-1}{\rm e}^{-\beta E_1}\delta_{\sigma,-\sigma}\{F(j'+\nu)[g_{01}(j)g_{01}(j') \nonumber\\
&&\quad+g_{12}(j)g_{12}(j+\nu)-g_{01}(j')g_{12}(j+\nu)]\nonumber\\
&&\quad+F(j)[g_{01}(j'+\nu)g_{01}(j+\nu)+g_{12}(j'+\nu)g_{12}(j') \nonumber\\
&&\quad-g_{01}(j+\nu)g_{12}(j')]\}, \label{C2}
\end{eqnarray}
where $E_0=0$, $E_1=-\mu$, and $E_2=U-2\mu$ are energies of the empty, singly, and doubly occupied states of the Hubbard atom with the Hamiltonian
$$H_{\bf l}=\sum_\sigma\bigg(\frac{U}{2}n_{\bf l\sigma}n_{\bf l,-\sigma}-\mu n_{\bf l\sigma}\bigg),$$
$g_{ii'}(j)=g_{ii'}(\omega_j)=({\rm i}\omega_j+E_i-E_{i'})^{-1}$, the atomic partition function $Z=\exp(-\beta E_0)+2\exp(-\beta E_1)+\exp(-\beta E_2)$, and $F(j)=g_{01}(j)- g_{12}(j)$. Equation~(\ref{C2}) differs from analogous expressions in Refs.~\citen{Vladimir,Pairault,Sherman06} in the quantity $\beta'=(T+\zeta)^{-1}$. For $\zeta=0$, as in these expressions, the vertex (\ref{Vs}) diverges at a finite temperature.\cite{Sherman18} This divergence signals the transition to the LRAFO and the finiteness of the transition temperature points to the violation of the Mermin-Wagner theorem. The introduction of a positive quantity $\zeta$, which somewhat suppresses spin fluctuations, is aimed to remedy this defect.\cite{Sherman19} Below, we discuss a method for the $\zeta$ determination.

Taking into account the contribution of ladder diagrams to the spin susceptibility (\ref{chisp}), it reads\cite{Sherman19}
\begin{eqnarray}
&&\chi^{\rm sp}({\bf q},\nu)=-\frac{T}{N}\sum_{{\bf k}j}G({\bf k},j)G({\bf k+q},j+\nu)\nonumber\\
&&\quad-T^2\sum_{jj'}F_{\bf q}(j,j+\nu)F_{\bf q}(j',j'+\nu)\nonumber\\
&&\quad\times V^s(j+\nu,j'+\nu,j',j), \label{sp}
\end{eqnarray}
where $F_{\bf q}(j,j')=N^{-1}\sum_{\bf k}\Pi({\bf k},j)\Pi({\bf k+q},j')$ and the terminal line $\Pi({\bf k},j)=1+t_{\bf k}G({\bf k},j)$. It describes the sum of all possible combinations of hopping lines and two-leg irreducible diagrams attached to free endpoints of extreme cumulants in ladders. Diagrams contributing to (\ref{sp}) are shown in Fig.~\ref{Fig1}(c).

In the considered set of diagrams, the superconducting susceptibility (\ref{chisc}) reads\cite{Sherman21}
\begin{eqnarray}
&&\chi^{\rm sc}=\frac{T}{N}\sum_p|\phi_{\bf k}|G(p)G(-p)\nonumber\\
&&\quad+\frac{T^2}{2N^2}\sum_{pp'}\phi_{\bf k'}\phi^*_{\bf k}\Pi(p)\Pi(-p)\Pi(p')\Pi(-p')\nonumber\\
&&\quad\times (W^+_{p'p}+W^-_{p'p}), \label{sc}
\end{eqnarray}
where, for brevity, we use the index $p$ combining the momentum ${\bf k}$ and the Matsubara frequency $\omega_j$. Quantities $W^+$ and $W^-$ are infinite sums of particle-particle reducible diagrams corresponding to singlet and triplet pairing, respectively. They satisfy the following BSE:
\begin{equation}
W^\pm_{p'p}=V^\pm_{p'p}+\frac{T}{2N}\sum_{p''}V^\pm_{p'p''}\theta(p'')\theta(-p'')
W^\pm_{p''p},\label{wpm}
\end{equation}
with
\begin{eqnarray}
&&V^+_{p'p}=-\frac{1}{2}[V^c_{\bf k'-k}(j',1-j,1-j',j)\nonumber\\
&&\quad+V^c_{\bf -k'-k}(1-j',1-j,j',j)] \nonumber\\
&&\quad+\frac{3}{2}[V^s_{\bf k'-k}(j',1-j,1-j',j)\nonumber\\
&&\quad+V^s_{\bf -k'-k}(1-j',1-j,j',j)] \nonumber\\
&&\quad-C^{(2a)}(j',1-j,1-j',j)\nonumber\\
&&\quad-C^{(2a)}(1-j',1-j,j',j),\nonumber\\[-1ex]
&&\label{Vpm}\\[-1ex]
&&V^-_{p'p}=\frac{1}{2}[-V^c_{\bf k'-k}(j',1-j,1-j',j)\nonumber\\
&&\quad+V^c_{\bf -k'-k}(1-j',1-j,j',j)\nonumber\\
&&\quad-V^s_{\bf k'-k}(j',1-j,1-j',j)\nonumber\\
&&\quad+V^s_{\bf -k'-k}(1-j',1-j,j',j)] \nonumber\\
&&\quad+C^{2a}(j',1-j,1-j',j)\nonumber\\
&&\quad-C^{2a}(1-j',1-j,j',j).\nonumber
\end{eqnarray}
Diagrams corresponding to Eqs.~(\ref{sc}) and (\ref{wpm}) are shown in Figs.~\ref{Fig1}(d) and (e). Notice that, in contrast to the case of the spin susceptibility, $V^s$ and $V^c$ contributing to $V^\pm$ are sums of vertical ladders.

The evidence of the superconducting transition is the divergence of one of the quantities $W^+$ or $W^-$ in the susceptibility (\ref{sc}). As follows from (\ref{wpm}), such a divergence takes place if the eigenvalue $E$ of the Eliashberg equation
\begin{equation}\label{Eliashberg}
\frac{T}{2N}\sum_{p}V^\pm_{p'p}\theta(p)\theta(-p)\varphi_p=E\varphi_{p'}
\end{equation}
attains unity.

A more detailed derivation of the above equations can be found in Refs.~\citen{Sherman18,Sherman19,Sherman21} and references therein.

Equations (\ref{Larkin})--(\ref{C2}) allow one to calculate the electron Green's function for given values of $\mu$, $T$, $U$ and hopping constants. The calculation is performed by iteration. As the initial irreducible part $K({\bf k},j)$, the first term on the right of Eq.~(\ref{K}) -- $C^{(1)}(j)$ -- is used. Green's function with this $K$ coincides with the result of the Hubbard-I approximation. To attain low temperatures, we have to introduce the parameter $\zeta$ into $C^{(2)}$ (\ref{C2}) and choose its value such that the transition to the LRAFO does not occur at a finite temperature. In Ref.~\citen{Sherman19}, $\zeta$ was fitted such that the transition exhibits at $T=0$ for half-filling when the spin fluctuations are the strongest. However, this value may not be appropriate for other fillings and temperatures.

In this work, as a criterion for evaluating $\zeta$, we use the equality of values of the squared site spin $\langle{\bf S}^2_{\bf l}\rangle$ obtained by two different methods -- from the one-electron Green's function\cite{Vilk} and from the spin susceptibility,
\begin{equation}\label{zeta}
\frac{\bar{n}}{2}-\frac{T}{UN}\sum_{{\bf k}j}{\rm e}^{{\rm i}\omega_j\eta}G({\bf k},j) \Sigma({\bf k},j)
=\frac{T}{N}\sum_{{\bf q}\nu}\chi^{\rm sp}({\bf q},\nu),
\end{equation}
where $\Sigma({\bf k},j)={\rm i}\omega_j-t_{\bf k}+\mu-[G({\bf k},j)]^{-1}$ is the self-energy, $\bar{n}=2TN^{-1}\sum_{{\bf k}j}\exp{({\rm i}\omega_j\eta)}G({\bf k},j)$
the electron concentration, and $\eta\rightarrow+0$. Equation~(\ref{zeta}) is convenient for estimating $\zeta$ since its left-hand side depends only weakly on this parameter, while the right-hand side significantly varies with $\zeta$ due to the staggered susceptibility $\chi^{\rm sp}({\bf Q},0)$ (${\bf Q}=(\pi,\pi)$, the intersite distance is set as the unit of length). Hence Green's function calculated with $\zeta$ from Ref.~\citen{Sherman19} can be used in the left-hand side of (\ref{zeta}), and this parameter in the right-hand side is fitted to fulfill the equation. The accuracy of such obtained parameter can be controlled by inserting Green's function calculated with the new value of $\zeta$ in the left-hand side of (\ref{zeta}). In contrast to the method of Ref.~\citen{Sherman19}, such obtained $\zeta$ depends on the temperature and filling. As will be seen in the next section, the magnetic susceptibility calculated with this parameter and quantities derived from it are in good agreement with results of the numeric and optical-lattice experiments and have a correct asymptotic behavior.

In calculations, we considered the cases of moderate and strong repulsions, $4\leq U/t\leq 8$ and the range of the chemical potential $T\ll\mu$, $T\ll U-\mu$. This range covers cases of half-filling, $\mu=U/2$, $T\ll U$, and moderate doping. In this range, expressions for cumulants (\ref{C1}), (\ref{C2}) are considerably simplified -- terms containing Boltzmann factors $\exp{(-\beta E_0)}$ and $\exp{(-\beta E_2)}$ can be omitted, and BSEs for $V^s$, Eq.~(\ref{Vs}), and $V^c$ are reduced to systems of four linear equations, which are easily solved.\cite{Sherman18} Outside of this range of chemical potentials, for $|\mu|\lesssim T$ and $|U-\mu|\lesssim T$, regions of the negative charge compressibility are located.\cite{Sherman20} This peculiarity of the model can lead to the formation of stripes. By choosing the mentioned range of $\mu$, we exclude them from consideration.

\section{Magnetic properties}
\begin{figure}[t]
\vspace*{3ex}
\centering
\includegraphics[width=0.99\columnwidth]{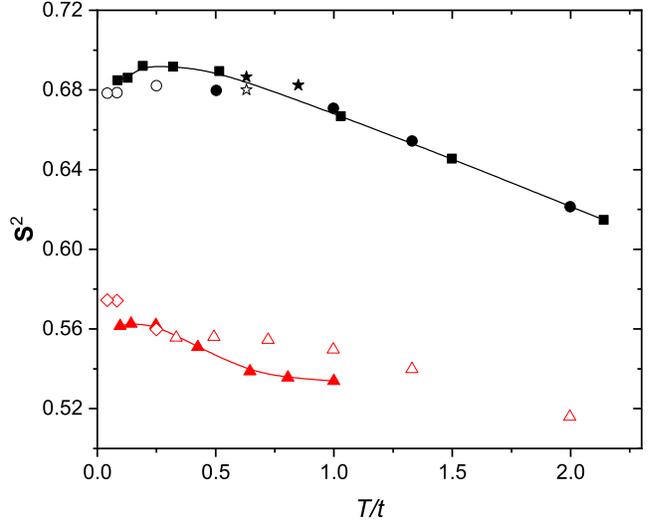}
\caption{(Color online) The temperature dependence of the squared site spin in the $t$-$U$ model for half-filling and two values of the Hubbard repulsion -- $U=8t$ (black symbols and line) and $U=4t$ (red symbols and line). Results of calculations using equations of the previous section are shown by filled squares and triangles. Data obtained by Monte Carlo simulations in a 6$\times$6 lattice are displayed by filled circles and open triangles,\protect\cite{Hirsch} and in a 10$\times$10 lattice by open circles and rhombuses.\protect\cite{Varney} Results of experiments with ultracold atoms\protect\cite{Drewes} and NLCE calculations\protect\cite{Khatami} are shown by filled and open stars, respectively.} \label{Fig2}
\end{figure}

Since the squared site spin is used in evaluating $\zeta$, we start from this quantity. Its value calculated from the spin susceptibility with $\zeta$ obtained from Eq.~(\ref{zeta}) is compared with results of the numeric and optical-lattice experiments in Fig~\ref{Fig2}.
As seen from the figure, the deviation of our calculated results from outcomes of the numeric and optical-lattice experiments is less than 2\% in a wide range of temperatures and for two different values of $U$.

\begin{figure}[t]
\vspace*{3ex}
\centering
\includegraphics[width=0.99\columnwidth]{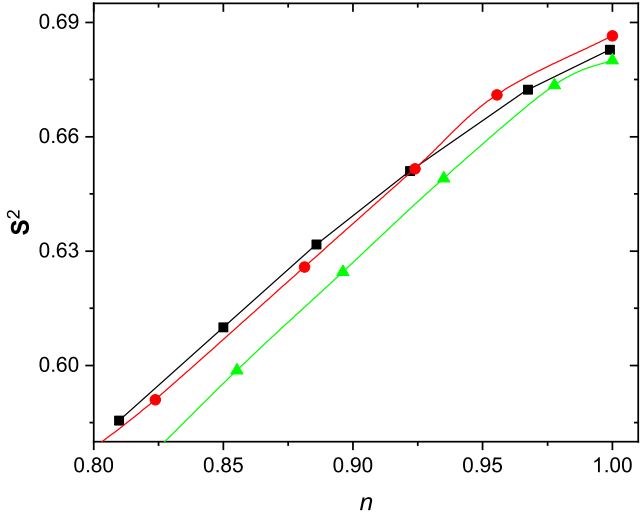}
\caption{(Color online) The concentration dependence of the squared site spin in the $t$-$U$ model. Black squares show results of SCDT calculations for $U=8t$ and $T=0.64t$. Red circles and green triangles are data of the experiment in a 2D optical lattice and NLCE calculations performed for $U=8.2t$ and $T=0.63t$, respectively.\protect\cite{Drewes}} \label{Fig3}
\end{figure}

In Fig.~\ref{Fig3}, the concentration dependence of $\langle{\bf S}_{\bf l}^2\rangle$ calculated by SCDT is compared with results of optical-lattice experiments and NLCE calculations. Again, one can conclude that the method of $\zeta$ calculation based on equation~(\ref{zeta}) gives a satisfactory result.

\begin{figure}[htb]
\centering
\includegraphics[width=0.99\columnwidth]{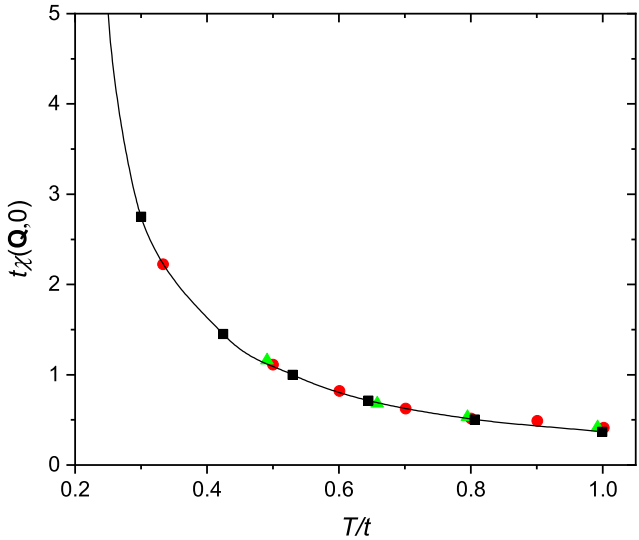}
\caption{(Color online) The temperature dependence of the staggered spin susceptibility in the $t$-$U$ model for half-filling and $U=4t$. The SCDT results are shown by filled black squares. Red circles and green triangles are Monte Carlo data obtained in 4$\times$4\cite{Bickers} and 6$\times$6\cite{Hirsch} lattices.} \label{Fig4}
\end{figure}

The temperature dependence of the staggered spin susceptibility $\chi^{\rm sp}({\bf Q},0)$ calculated with $\zeta$ from Eq.~(\ref{zeta}) is shown in Fig.~\ref{Fig4}. The agreement with the results of Monte Carlo simulations is satisfactory and much better than that achieved with the zero-temperature value $\zeta=0.24t$ in Fig.~3 in Ref.~\citen{Sherman19}. For low but finite temperatures, the parameter $\zeta$ is smaller than this value, which leads to a greater staggered susceptibility $\chi^{\rm sp}({\bf Q},0)$. It is the reason for the rapid temperature growth of the $d_{x^2-y^2}$ eigenvalue of the Eliashberg equation in the $t$-$t'$-$t''$-$U$ model.

\begin{figure}[t]
\vspace*{2ex}
\centering
\includegraphics[width=0.99\columnwidth]{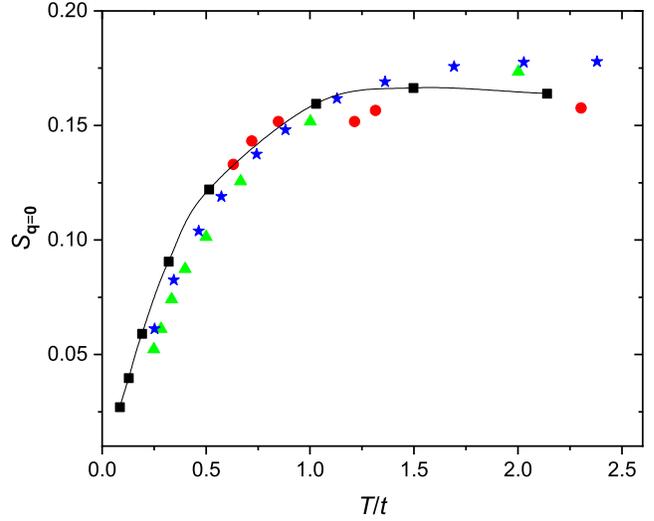}
\caption{(Color online) The temperature dependence of the spin structure factor in the $t$-$U$ model at half-filling. Black squares, blue stars, and green triangles are results of SCDT, NLCE,\protect\cite{Khatami} and Monte Carlo simulations\protect\cite{Paiva} for $U=8t$, respectively. Red circles are data of experiments in 2D optical lattices measured at $U=8.2t$.\protect\cite{Drewes}} \label{Fig5}
\end{figure}

Figure~\ref{Fig5} demonstrates the temperature behavior of the spin structure factor
\begin{eqnarray}\label{S_0}
S_{\bf q=0}&=&\frac{1}{2}\sum_{\bf l}\langle s^+_{\bf l}s^-_{\bf 0}\rangle=\frac{T}{2}\sum_\nu\chi^{\rm sp}({\bf 0},\nu)\nonumber\\
&=&\frac{T}{2}\chi^{\rm sp}({\bf 0},0).
\end{eqnarray}
The last equality in Eq.~(\ref{S_0}) follows from the fact that $\chi^{\rm sp}({\bf 0},\nu)=0$ for all $\nu\neq0$.\cite{Vilk} Thus, the spin structure factor is determined by the uniform susceptibility. As seen from Fig.~\ref{Fig5}, our results with $\zeta$ defined by Eq.~(\ref{zeta}) are in satisfactory agreement with outcomes of Monte Carlo, NLCE calculations, and experiments in optical lattices. The achieved agreement is much better than that in Fig.~9 of Ref.~\citen{Sherman19} obtained with the zero-temperature $\zeta$.

Another verification of the used way for determining $\zeta$ can be obtained from the comparison of the asymptotic behavior of the spin susceptibility with the frequency dependence following from the electron Green's function. From spectral representations and the symmetry of the Hamiltonian, one can show that $\chi^{\rm sp}({\bf q},\nu)$ is an even function of the Matsubara frequency $\omega_\nu=2\pi\nu T$ with real and positive values. Using the spectral representations and calculating commutators of spin components with the Hamiltonian (\ref{Hamiltonian}), one can find
\begin{equation}\label{asymptote}
\chi^{\rm sp}({\bf q},\nu)\stackrel{|\nu|\rightarrow\infty}{\longrightarrow} \frac{2}{N\omega_\nu^2}\sum_{\bf k}t_{\bf k}(c_{\bf q+k}-c_{\bf k}),
\end{equation}
where
$$c_{\bf k}=\sum_{\bf l}{\rm e}^{{\rm i}{\bf k}({\bf l-l'})}\big\langle a^\dagger_{\bf l}a_{\bf l'}\big\rangle=T\sum_j{\rm e}^{{\rm i}\omega_j\eta}G({\bf k},j)$$
with $\eta\rightarrow+0$. For the $t$-$U$ model, Eq.~(\ref{asymptote}) is simplified to
\begin{equation}\label{asymp_tU}
\chi^{\rm sp}({\bf q},\nu)\stackrel{|\nu|\rightarrow\infty}{\longrightarrow} \frac{8tc_1(1-\gamma_{\bf q})}{\omega_\nu^2},
\end{equation}
where
$$c_1=\frac{T}{N}\sum_{{\bf k}j}{\rm e}^{{\rm i}\omega_j\eta}\gamma_{\bf k}G({\bf k},j),\;\gamma_{\bf k}=\frac{1}{2}[\cos(k_x)+\cos(k_y)].$$
\begin{figure}[t]
\vspace*{3ex}
\centering
\includegraphics[width=0.99\columnwidth]{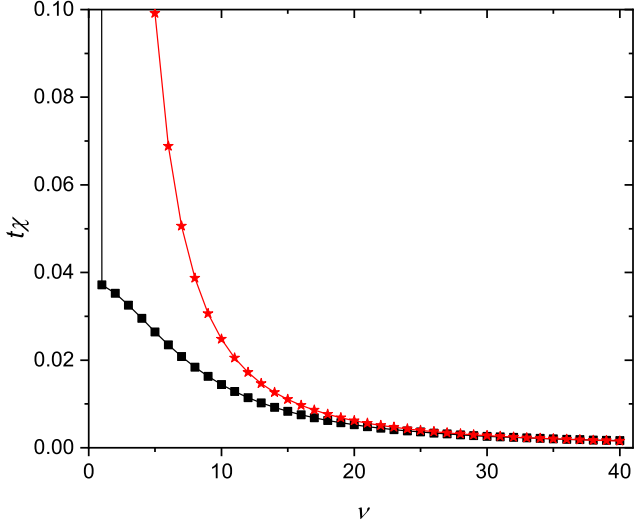}
\caption{(Color online) The calculated dependence of the spin susceptibility on the imaginary frequency $\omega_\nu=2\pi\nu T$ (black squares) is compared with the asymptotics (\protect\ref{asymp_tU}) (red stars). Calculations were performed for the $t$-$U$ model, $U=8t$, $T\approx0.13t$, and ${\bf q=Q}$.} \label{Fig6}
\end{figure}
In Fig.~\ref{Fig6}, we compared the calculated spin susceptibility with $\zeta$ determined from Eq.~(\ref{zeta}) with the asymptotics (\ref{asymp_tU}). As seen from the figure, two dependencies coalesce for $\nu\gtrsim 20$. Similar results are obtained for other momenta, values of $U$, and in the $t$-$t'$-$t''$-$U$ model. As mentioned above, $\chi^{\rm sp}({\bf 0},\nu)$ has to vanish for $\nu>0$. In our calculations, this result is obtained only approximately, with $\chi^{\rm sp}({\bf 0},\nu\neq0)\lesssim10^{-4}t^{-1}$.

It is instructive to compare Eq.~(\ref{asymp_tU}) with the expression for the spin susceptibility in the Heisenberg model with nearest-neighbor exchange interaction derived in Ref.~\citen{Shimahara}. For the 2D case, this expression reads
\begin{equation}\label{Shimahara}
\chi^{{\rm H}}({\bf q},\nu)=\frac{8J|C_1|(1-\gamma_{\bf q})}{\omega_\nu^2+\omega_{\bf q}^2},
\end{equation}
where $J$ is the exchange constant, $C_1=\langle s^+_{\bf l}s^-_{\bf l'}\rangle\approx-0.2$ for $T=0$ and nearest neighbor sites {\bf l} and ${\bf l'}$, $\omega_{\bf q}$ is the frequency of spin excitations. For the case of half-filling, the strong repulsion $U=8t$, and low temperatures $T\lesssim0.2t$, the quantity $c_1\approx0.1$. For this repulsion, $J=4t^2/U=0.5t$. Hence, as could be expected, the asymptotics of the susceptibility (\ref{Shimahara}) is close to that given by Eq.~(\ref{asymp_tU}).

\begin{figure}[htb]
\centering
\includegraphics[width=0.99\columnwidth]{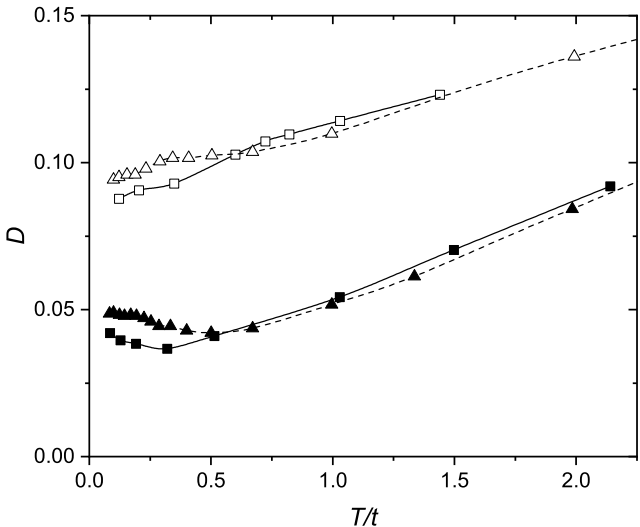}
\caption{(Color online) The temperature dependence of the double occupancy in the half-filled $t$-$U$ model calculated for $U=8t$ (filled squares) and $U=5.1t$ (open squares). For comparison, results of Monte Carlo simulations\protect\cite{Paiva} for $U=8t$ (filled triangles) and $U=5t$ (open triangles) are also shown.} \label{Fig7}
\end{figure}
The above results show that the case of strong repulsions and low temperatures is characterized by well-defined local spins. Indeed, in this case, the squared site spin is close to its maximum value of $3/4$, and the asymptotic behavior of the spin susceptibility is close to that in the Heisenberg model. Another indicator of local spins is a small value of the double site occupancy $D=\langle n_{\bf l\uparrow}n_{\bf l\downarrow}\rangle$. Results of calculations of this quantity in the half-filled $t$-$U$ model are given in Fig.~\ref{Fig7} for moderate and strong Hubbard repulsions. For comparison, data of Monte Carlo simulations \cite{Paiva} are also shown in this figure. The agreement is satisfactory.

Summarizing, we can notice that the precise determination of the parameter $\zeta$ allows us to obtain a much better agreement of calculated quantities with data of numeric and optical-lattice experiments.

\section{Superconductivity}
As mentioned in the previous section, for low temperatures, values of the parameter $\zeta$ determined by Eq~(\ref{zeta}) are smaller than those found for $T=0$. As a consequence, the spin vertex $V^s$ becomes larger. It enters into the matrix $M_{p'p}=(T/2N)V^\pm_{p'p}\theta(p)\theta(-p)$ of the Eliashberg equation (\ref{Eliashberg}). Hence the increase of this vertex may substantially enhance eigenvalues of the matrix in comparison with our previous calculations.\cite{Sherman21} Therefore, in this section, we recalculate the eigenvalues with newly defined $\zeta$.
\begin{figure}[t]
\vspace*{3ex}
\centering
\includegraphics[width=0.99\columnwidth]{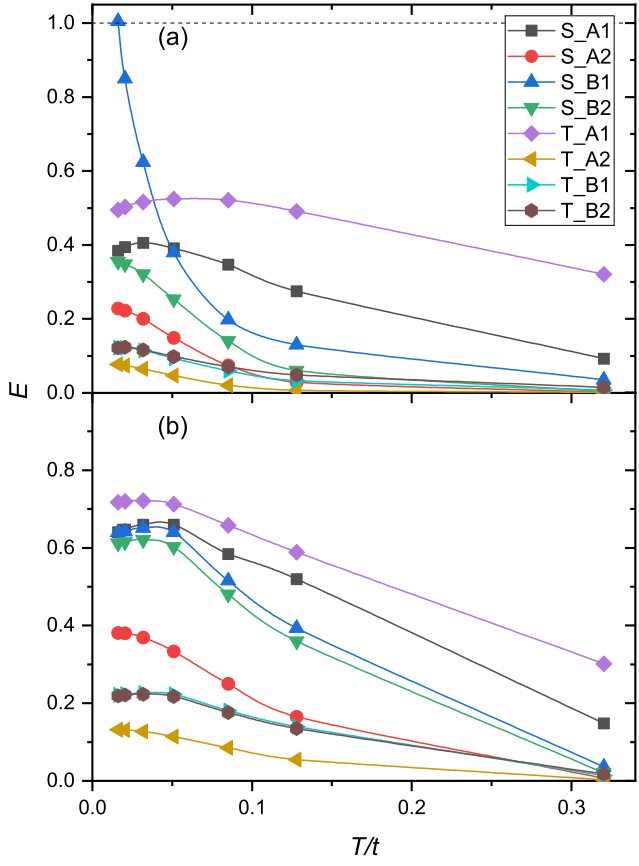}
\caption{(Color online) Eigenvalues of the Eliashberg equation (\protect\ref{Eliashberg}) as functions of temperature for $U=8t$, $t_1=-0.3t$, $t_2=0.2t$ (a) and $U=8t$, $t_1=t_2=0$ (b). For both models $\bar{n}=0.92$. Lines and symbols of different colors correspond to different pairing symmetries indicated in the legend, in which letter S points to singlet and T to triplet pairing.} \label{Fig8}
\end{figure}

In these calculations, both the $t$-$U$ and $t$-$t'$-$t''$-$U$ models are considered at the chemical potential $\mu\approx2t$. For the two models, with different values of $U$ and $T$, the chemical potential was slightly varied around this value to attain the electron concentration $\bar{n}=0.92$ for all considered sets of parameters. Such a choice of $\mu$, on the one hand, is connected with the mentioned simplification of BSEs for $V^s$ and $V^c$ in the range of chemical potentials $T\ll\mu$, $T\ll U-\mu$. On the other hand, such a value of $\mu$ allows us to avoid the regions of the negative electron compressibility located near $\mu=0$ and $\mu=U$. \cite{Sherman20} In calculations including phonons or with an unfixed chemical potential, these regions lead to charge instability and phase separation.

The matrix $M_{p'p}$ is invariant to the transformations of the point group $D_4$ of the lattice. Therefore, its eigenvectors are characterized by representations of this group. There are five such representations, four of which -- $A_1$ ($x^2+y^2$), $A_2$ ($z$), $B_1$ ($x^2-y^2$), and $B_2$ ($xy$) -- are one-dimensional and one is two-dimensional.\cite{Bir} We limit ourselves to the one-dimensional representations.

To solve the Eliashberg equation (\ref{Eliashberg}), we use the power (von Mises) iteration.\cite{Mises} The largest eigenvalues for singlet and triplet pairings and all one-dimensional representations are shown in Fig~\ref{Fig8}. Comparing it with Fig.~2 in Ref.~\citen{Sherman21}, we see that the eigenvalues have significantly increased. Nevertheless, for the $t$-$U$ model, they are still less than unity (see panel (b)), which signals the absence of the superconducting transitions. However, for the $t$-$t'$-$t''$-$U$ model with the mentioned hopping constants, the eigenvalue of the singlet $B_1$ ($d_{x^2-y^2}$) solution attains unity at $T_c\approx0.016t$ (see panel (a)). Hence the model exhibits the transition to the superconducting state.
\begin{figure}[t]
\vspace*{3ex}
\centering
\includegraphics[width=0.99\columnwidth]{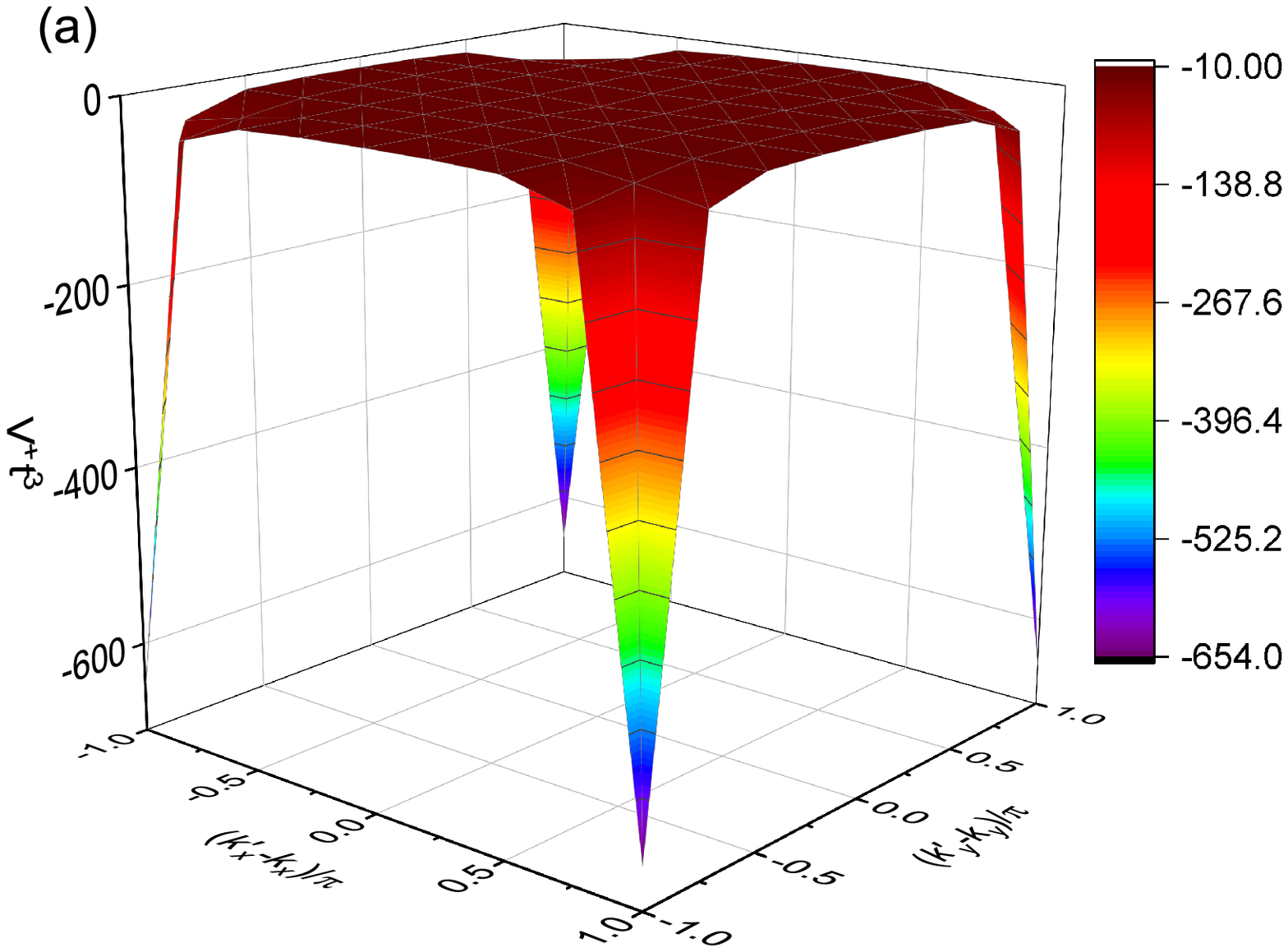}

\vspace*{2ex}
\includegraphics[width=0.99\columnwidth]{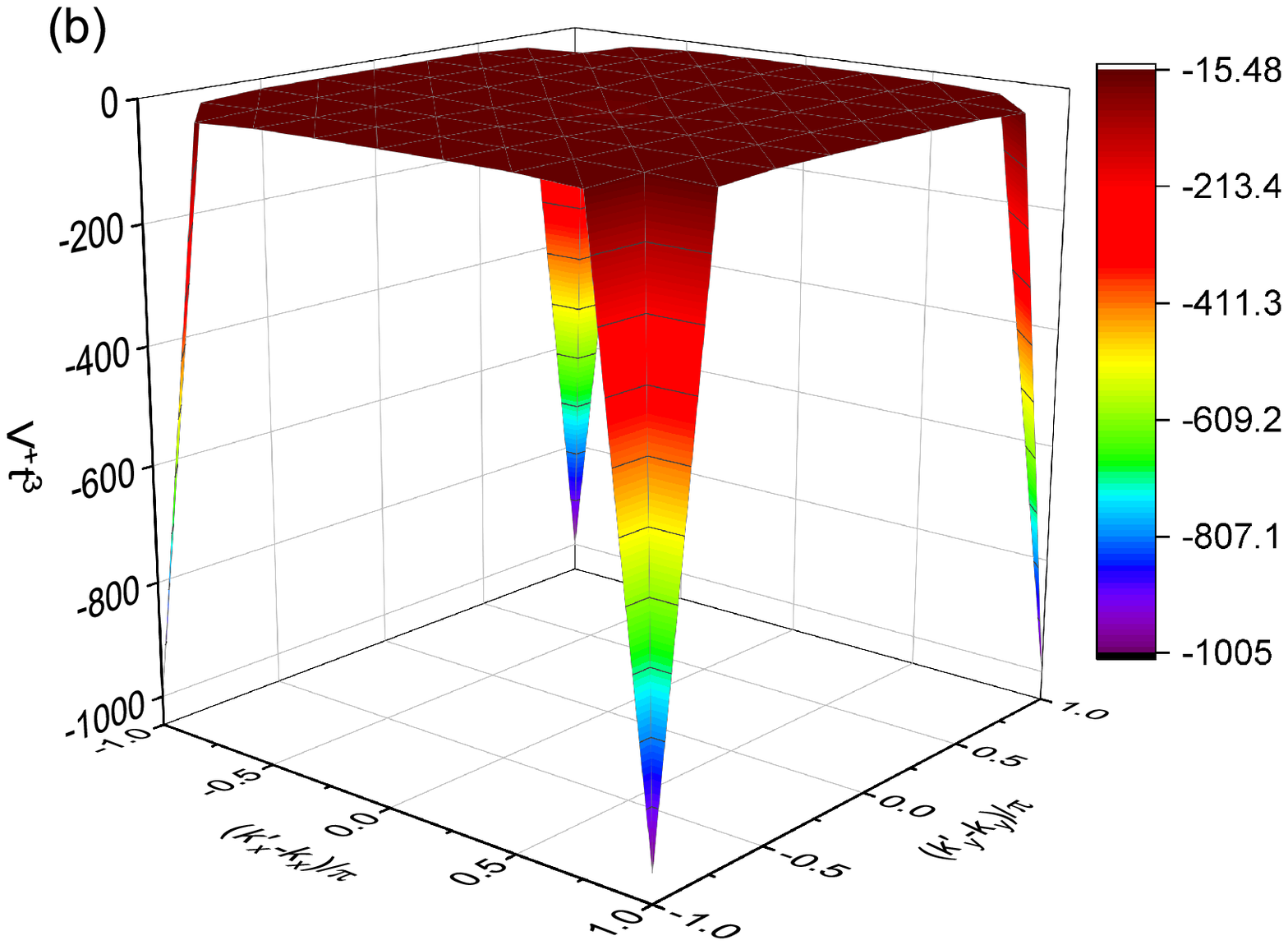}
\caption{(Color online) The momentum dependence of $V^+_{\bf k'-k}(j',j)$, Eq.~(\protect\ref{Vpm}), for $U=8t$, $t_1=-0.3t$, $t_2=0.2t$ (a) and $U=8t$, $t_1=t_2=0$ (b). In both cases, $T\approx0.02t$, $\bar{n}\approx0.92$, and $j'=j=0$.} \label{Fig9}
\end{figure}

\begin{figure}[t]
\vspace*{2ex}
\centering
\includegraphics[width=0.99\columnwidth]{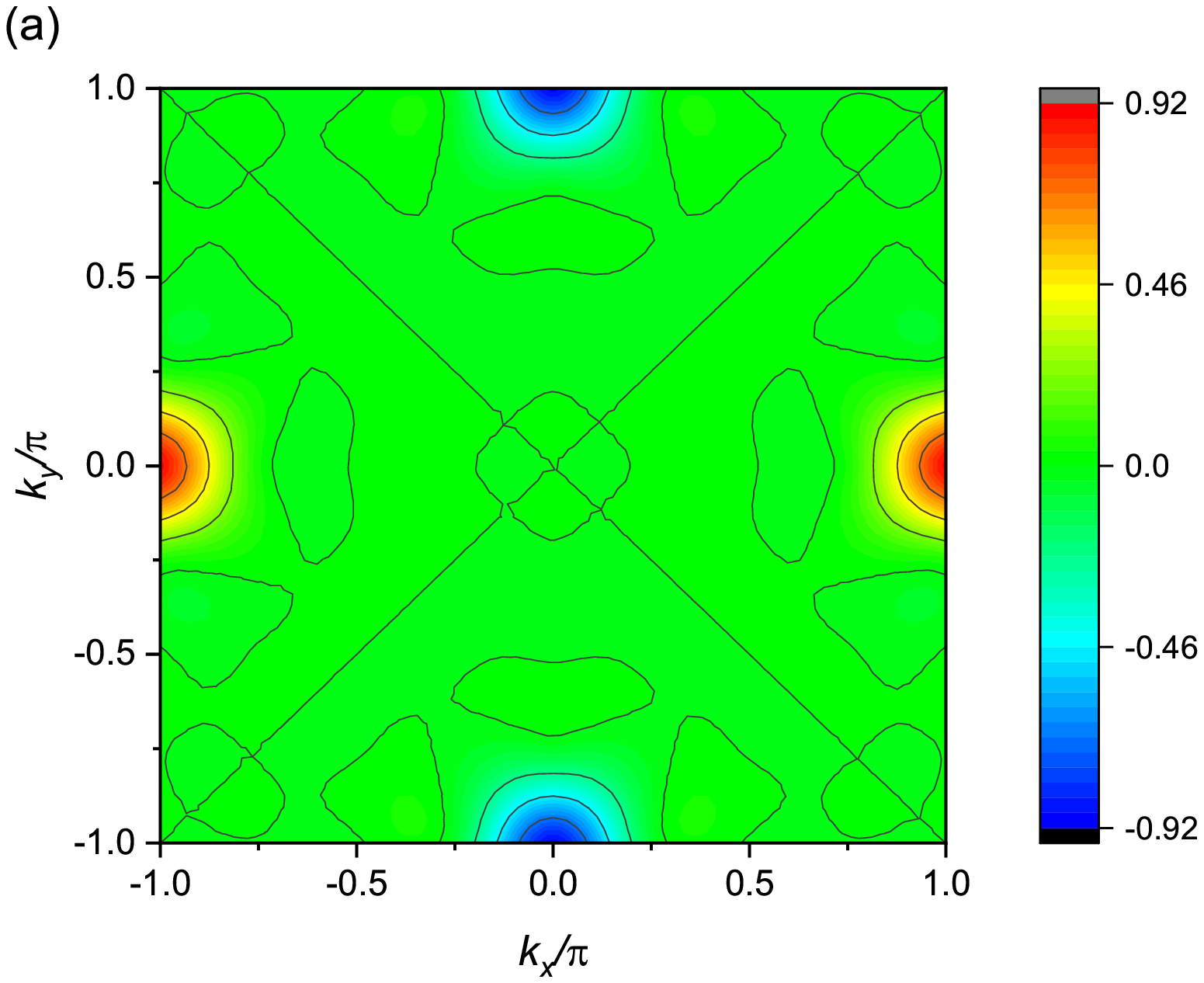}
\includegraphics[width=0.99\columnwidth]{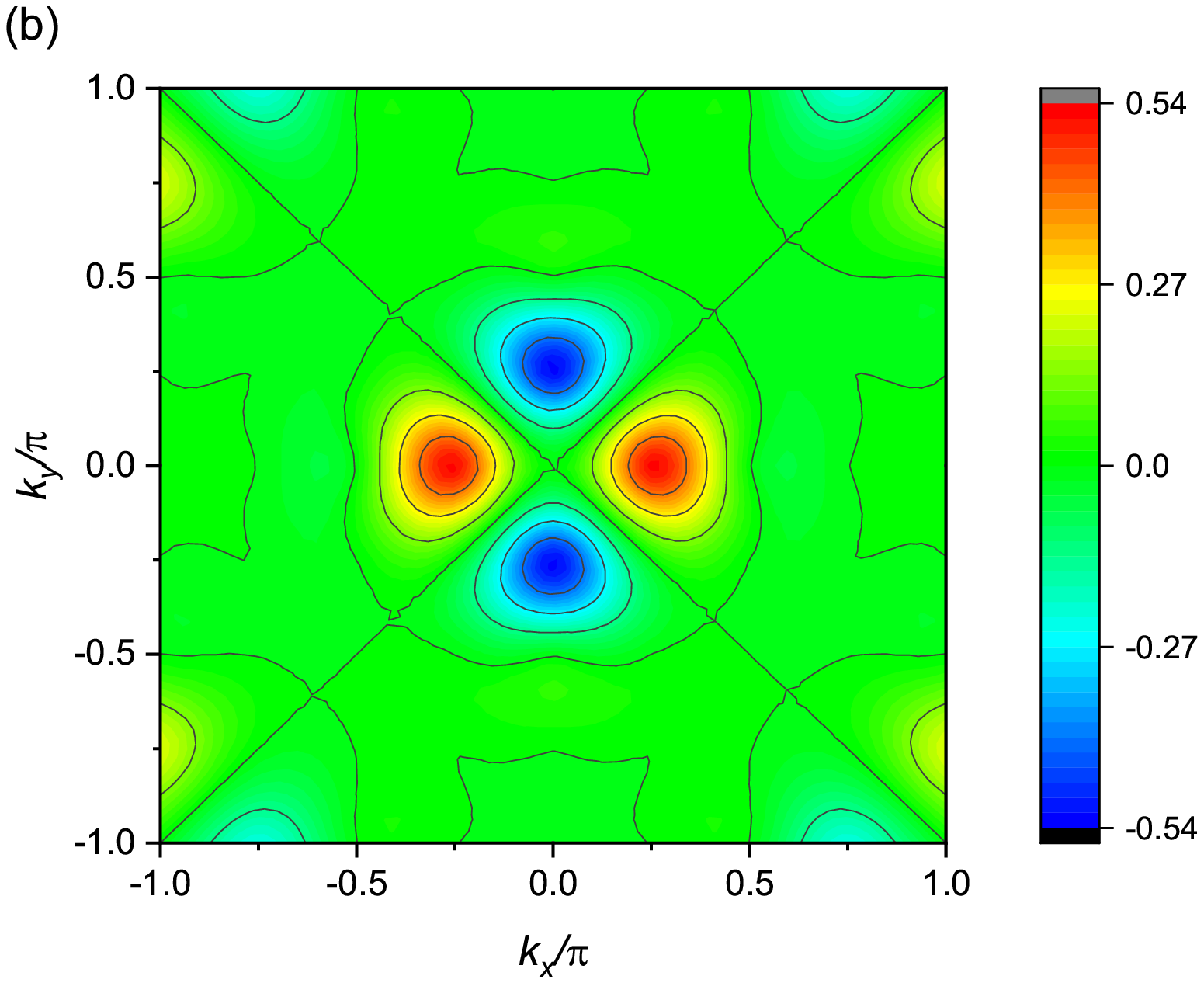}
\caption{(Color online) Contour plots of the momentum dependence of singlet $B_1$ eigenvectors related to the largest eigenvalues of the Eliashberg equation (\protect\ref{Eliashberg}) for $U=8t$, $t_1=-0.3t$, $t_2=0.2t$ (a) and $U=8t$, $t_1=t_2=0$ (b). In both cases, $T\approx0.02t$, $\bar{n}\approx0.92$. The parameter $j$ defining the Matsubara frequency $\omega_j=(2j-1)\pi T$ is equal to $-9$ in panel (a) and 0 in panel (b). These $j$ correspond to the largest in absolute value components of the respective eigenvectors.}  \label{Fig10}
\end{figure}

What is the reason for this difference in the behavior of the two models? The sums of ladder diagrams $V^+$ entering into the matrix of the Eliashberg equation are shown in Fig.~\ref{Fig9} for both models. These quantities are real. Here we allowed for that solutions corresponding to one-dimensional representations of the $D_4$ group are invariant under inversion. For them, momenta ${\bf -k'}$ in spin and charge vertices $V^s$ and $V^c$ in Eq.~(\ref{Vpm}) can be substituted with ${\bf k'}$. As a result, $V^+$ depends only on three variables -- ${\bf k'-k}$, $j'$, and $j$. Quantities $V^+$ for both models have sharp minima at ${\bf k'-k}={\bf Q}$, which indicates that pronounced antiferromagnetic fluctuations described by $V^s$ make the main contribution to this minima. Therefore, singlet $B_1$ eigenvectors corresponding to the largest eigenvalues of the Eliashberg equation,
\begin{equation}\label{eigen}
E=\frac{\sum_{{\bf k'}j'}\sum_{{\bf k}j}\varphi^*_{\bf k'}(j')M_{\bf k'-k}(j',j)\varphi_{\bf k}(j)}{\sum_{{\bf k}j}\varphi^*_{\bf k}(j)\varphi_{\bf k}(j)},
\end{equation}
should have large in absolute values and opposite in sign components $\varphi^*_{\bf k'}(j')$, $\varphi_{\bf k}(j)$ for momenta ${\bf k'}$ and ${\bf k}$ satisfying the relation ${\bf k'-k}={\bf Q}$. Regions of such components in the Brillouin zone are seen in Fig.~\ref{Fig10}, which shows contour plots of the eigenvectors in the considered models.

Two peculiarities attract the attention in Figs.~\ref{Fig9} and \ref{Fig10}. First, in spite that the singlet $B_1$ eigenvalue is larger for $t_1=-0.3t$, $t_2=0.2t$ than for $t_1=t_2=0$, minima of $V^+$ are deeper in the latter case than in the former. Such differences in these quantities would be expected since the electron hopping to sites of second and third neighbors introduces frustration in the spin subsystem. Second, in spite that both contour plots in Fig.~\ref{Fig10} have distinctive features of the $d_{x^2-y^2}$ symmetry, they are drastically different. It is clear that the cause of these peculiarities is in the two multipliers $\theta(p)$ and $\theta(-p)$, which together with $V^+_{p'p}$ form the matrix in the Eliashberg equation (\ref{Eliashberg}).

\begin{figure}[t]
\vspace{2ex}
\centering
\includegraphics[width=0.99\columnwidth]{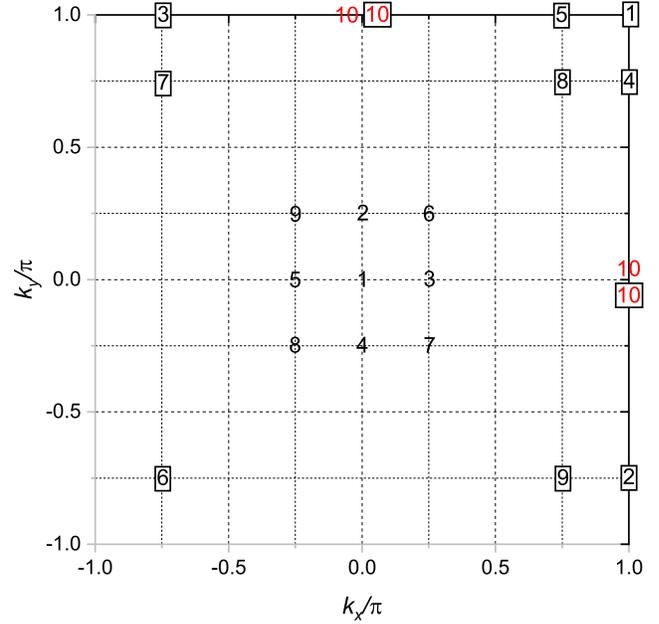}
\caption{(Color online) Momenta of the largest in absolute value components of the matrix $V^+_{\bf k'-k}(j',j)\theta({\bf k},j)\theta({\bf -k},1-j)$ in the Eliashberg equation (\protect\ref{Eliashberg}) in an 8$\times$8 lattice. Vectors ${\bf k'}$ are indicated by numbers, the respective vectors ${\bf k}$ by the same numbers in rectangles. Vector pairs shown by black numbers occur both in the $t$-$U$ and $t$-$t'$-$t''$-$U$ models, while those indicated by red numbers only in the latter. $U=8t$, $T\approx0.02t$, and $\bar{n}\approx0.92$ in both models. In the $t$-$t'$-$t''$-$U$ model, $t_1=-0.3t$, $t_2=0.2t$.}  \label{Fig11}
\end{figure}
Figure~\ref{Fig11} shows the location of wave vectors of the hundred largest in absolute value components of the matrix $V^+_{\bf k'-k}(j',j)\theta({\bf k},j)\theta({\bf -k},1-j)$ in the Eliashberg equation (\ref{Eliashberg}) in an 8$\times$8 lattice. Parameters are the same as in the previous figures. Vectors ${\bf k'}$ are indicated by numbers in the place of their heads in the Brillouin zone. The respective vectors ${\bf k}$ are shown by the same numbers in rectangles. As indicated, the difference between these vectors is $(\pm\pi,\pm\pi)$. The number of pairs ${\bf k'}$ and ${\bf k}$ is modest since there are a lot of $j'$ and $j$ corresponding to each pair. Among the largest components, the number of pairs with momenta 2-\fbox{2}, \ldots 5-\fbox{5} is mich larger than pairs 1-\fbox{1}, 6-\fbox{6}, \ldots 9-\fbox{9}. The matrix components with momenta from the former set are responsible for the shape of the solution shown in Fig.~\ref{Fig10}(b). Vector pairs shown by black numbers occur both in the $t$-$U$ and $t$-$t'$-$t''$-$U$ models, while the pairs $(\pi,0)$-$(0,\pi)$ and $(0,\pi)$-$(\pi,0)$ indicated by red tens only in the latter model. The reason for this difference is that $\theta({\bf k},j)=t_{\bf k}=0$ at ${\bf k}=(\pi,0)$, $(0,\pi)$ for $t_1=t_2=0$. Evidently, the matrix extrema at the pairs of momenta $(\pi,0)$-$(0,\pi)$ and $(0,\pi)$-$(\pi,0)$ are responsible for the shape of the eigenvector in Fig.~\ref{Fig10}(a). This shape with the equally strong maximum and minimum ensures the large eigenvalue of the Eliashberg equation in the $t$-$t'$-$t''$-$U$ model.

As follows from the above discussion, there are opposite trends in multipliers $V^+_{\bf k'-k}(j',j)$ and $\theta({\bf k},j)\theta({\bf -k},1-j)$ in the matrix of the Eliashberg equation (\ref{Eliashberg}) -- a grows of $|t_1|$ and $t_2$ leads to an increase of the latter factor and decrease, due to frustration, of the former. Hence, for the one-band model, there exist optimal values of $t_1$ and $t_2$, which provide the maximum $T_c$. However, higher transition temperatures can be expected in multiband models. In this case, which is more flexible due to band hybridization, the competition of the spin vertex and hopping multipliers can be weakened. Using the value of the exchange constant $J=4t^2/U=0.1$~eV, as observed in cuprates, for $U=8t$, we find $t=0.2$~eV. Thus, the superconducting transition temperature found in our calculations $T_c\approx0.016t\approx37$~K, which is close to $T_c$ observed in La$_{2-x}$Ba$_x$CuO$_4$.

There are a lot of works devoted to the superconductivity in the repulsive Hubbard model (see, e.g., Refs.~\citen{Bickers,Yokoyama,Gros,Scalapino93,Bulut,Sorella,Maier,Senechal,Maier06,Capone,Aimi,Eichenberger,Aichhorn,Scalapino12,Gull,Misawa,Otsuki,Kitatani15,Vucicevic,Kitatani19,Qin}).
In the considered case of strong electron correlations, both affirmative and negative answers were obtained regarding the possibility of the superconducting transition. Let us contrast our obtained results with outcomes of some of these works, which investigated similar models. In Refs.~\citen{Maier06,Otsuki,Kitatani19}, Eliashberg equations similar to Eq.~(\ref{Eliashberg}) were derived using the weak coupling diagram technique. However, four-leg vertices analogous to $V^\pm$ were borrowed from a cluster diagonalization in Ref.~\citen{Maier06} and the Anderson impurity model in Refs.~\citen{Otsuki,Kitatani19}. In the Eliashberg equation (\ref{Eliashberg}) derived using the strong coupling expansion, the motion of the electron pair is described by the product $\theta(p)\theta(-p)$. In the equations derived in Refs.~\citen{Maier06,Otsuki,Kitatani19} with the weak coupling expansion, the same role is played by the product of two Green's functions $G(p)G(-p)$. Its value at ${\bf k}=(\pi,0)$, $(0,\pi)$ is nearly the same in the $t$-$U$ and $t$-$t'$-$t''$-$U$ models. Hence the difference in the character of superconducting fluctuations in these models is lost. As a result, in the $t$-$U$ model, the tendency toward the singlet $d$-wave superconductivity was found in Ref.~\citen{Maier06}. In the same model, the transition was observed in Refs.~\citen{Otsuki,Kitatani19}. In these works, the Hubbard repulsion was in the range from $4t$ to $8t$, the electron concentration was near the optimal doping $\bar{n}\approx0.85$, and the transition temperatures were of the order of $0.01t$. In Ref.~\citen{Kitatani19}, superconductivity with nearly the same $T_c$ was observed in the $t$-$t'$-$t''$-$U$ model also. The superconducting pairing correlations were studied using the auxiliary-field quantum Monte Carlo and density matrix renormalization group methods in Ref.~\citen{Qin}. The authors concluded that the ground state of the $t$-$U$ model is nonsuperconducting. The results were obtained near optimal doping in the presence of stripe order. For strong coupling ($U/t\approx6-8$), the authors supposed that the absence of superconductivity is the consequence of competition with these stripes. However, at smaller $U\approx4t$, for which the tendency of the stripe formation is much weaker, they still found a pairing response consistent with zero. This result is in accord with our conclusion that, even without static stripes, there is no superconducting transition in the $t$-$U$ model.

\section{Conclusion}
In this work, we introduced an improved method for evaluating the parameter $\zeta$ governing the strength of spin fluctuations and ensuring the fulfillment of the Mermin-Wagner theorem in the strong coupling diagram technique. The approach is based on the requirement on equality of squared site spin values obtained from one-particle and two-particle Green's functions. This condition uniquely defines the parameter for given values of $U$, $\mu$, $T$, and hopping constants. In previous works, $\zeta$ was determined at half-filling and $T=0$ and used for conditions differing from this situation. 

The aim of the new method for evaluating $\zeta$ was to improve the agreement of quantities derived from two-particle Green's functions with results of numeric and optical-lattice experiments.
Calculations carried out in this work showed that this agreement is indeed improved significantly with new values of $\zeta$. Among results used for comparison with experimental outcomes were the temperature and concentration dependencies of the squared site spin, the temperature dependencies of the staggered susceptibility, spin structure factor, and double occupancy. We also found that the frequency dependence of the spin susceptibility falls on the asymptotics defined by the electron Green's function.

With new $\zeta$, at low temperatures, the spin vertex considerably increases in comparison with its magnitude for old values of this parameter. To check whether this growth can lead to superconducting instability, we solved the Eliashberg equation both for the $t$-$U$ and $t$-$t'$-$t''$-$U$ models for the case of strong correlations, $U=8t$, and the electron concentration $\bar{n}=0.92$. This filling was chosen for two reasons. On the one hand, Bethe-Salpeter equations for spin and charge vertices are considerably simplified for chemical potentials in the range $T\ll\mu$, $T\ll U-\mu$. For considered temperatures, $\mu$ corresponding to the mentioned concentration falls into this range. On the other hand, this chemical potential is far enough from regions of the negative electron compressibility near $\mu=0$ and $\mu=U$, which influence we wish to eliminate.

In these calculations, both singlet and triplet pairing and order parameters belonging to all one-dimensional representations of the $D_4$ point group were considered. The investigation showed that the superconducting transition does not occur in the $t$-$U$ model. However, the $t$-$t'$-$t''$-$U$ model with parameters $t'=-0.3t$, $t''=0.2t$ exhibits the transition into the superconducting state with the singlet $B_1$ ($d_{x^2-y^2}$) pairing. The transition temperature $T_c\approx0.016t\approx37$~K.

The difference between the two models, which leads to the absence of superconductivity in one of them and its appearance in the other, is the value of the renormalized hopping $\theta({\bf k},j)$ at the momenta ${\bf k}=(\pi,0)$, $(0,\pi)$, that is the points of the maxima and minima of the $d$-wave order parameter. In the $t$-$U$ model, $\theta\propto[\cos(k_x)+\cos(k_y)]$ vanishes in these points, which decreases the eigenvalue of the Eliashberg equation. In the $t$-$t'$-$t''$-$U$ model, $\theta$ is finite in the points and makes a large contribution to the eigenvalue. Besides the product $\theta({\bf k},j)\theta({\bf -k},1-j)$ describing the motion of the electron pair, the matrix in the Eliashberg equation contains an additional multiplier -- the four-leg vertex. It has a sharp minimum at the corner of the Brillouin zone, which results from strong antiferromagnetic fluctuations described by the spin vertex. The magnitude of the minimum is another factor contributing to the eigenvalue. The growth of $|t_1|$ and $t_2$ leads to an increase in the product $\theta({\bf k},j)\theta({\bf -k},1-j)$ for ${\bf k}=(\pi,0)$, $(0,\pi)$. However, simultaneously, this growth decreases the magnitude of the vertex minimum due to the frustration in the spin subsystem introduced by electron hopping to second and third neighbors. These opposite trends in terms of the matrix define an optimal set of hopping constants producing the highest $T_c$ in the one-band Hubbard model.

\end{document}